\documentclass[
reprint,onecolumn,
superscriptaddress, 12pt, 
nofootinbib,
 amsmath,amssymb,
floatfix
]{revtex4-2}

\usepackage{physics} 
\usepackage{graphicx}
\usepackage{dcolumn}
\usepackage{bm}
\usepackage{cancel}
\usepackage{amsfonts}
\usepackage{xcolor}
\usepackage{tcolorbox}
\usepackage{placeins}
\usepackage{float}

\newcommand{\beqa}{\begin{eqnarray}}
\newcommand{\eeqa}{\end{eqnarray}}
\newcommand{\nn}{\nonumber}
\usepackage{dcolumn}
\usepackage{bm}

\usepackage{amsmath, fullpage,setspace}
\usepackage{graphicx}
\usepackage{subcaption}

\begin{document}

\title{Dissipative Vortex Binaries in Compact Fluid Domains with Geometric Corrections}

\author{Aswathy K.R.}
\affiliation{Birla Institute of Technology and Science, Pilani, Hyderabad Campus, Telangana 500078, India}

\author{Rickmoy Samanta}
\affiliation{Birla Institute of Technology and Science, Pilani, Hyderabad Campus, Telangana 500078, India}
\affiliation{Indian Institute of Technology Kharagpur, West Bengal 721302, India}

\begin{abstract}
We study a dissipative extension of vortex-binary motion in a doubly periodic fluid domain. The underlying conservative system admits an exact integrable reduction to a single complex relative coordinate. Dissipation is introduced via a minimal rotated-velocity (mutual-friction) term, as motivated by finite-temperature superfluid dynamics, converting the Hamiltonian evolution into a mixed symplectic-gradient flow with monotonic energy decay for quantized vortices. In the local regime, the dissipative binary remains analytically solvable and admits closed-form solutions, with systematic corrections arising from the toroidal geometry. Equal same-sign vortices execute outward spiraling motion, while equal opposite-sign pairs (dipoles) undergo finite-time collapse in the planar limit. On the torus, however, the dipole orientation is no longer invariant: the geometry induces a slow angular drift, even in regimes where planar dynamics would preserve alignment. For unequal opposite-sign pairs, dissipation induces coupled contraction and rotation, leading to a finite-time nonlinear chirp characterized by $\dot{\omega}\propto\omega^2$, in contrast with electromagnetic and gravitational inspirals where $\dot{\omega}\propto \omega^{3}$ and $\dot{\omega}\propto \omega^{11/3}$. These results highlight the interplay between Hamiltonian structure, dissipation, and geometry in periodic fluid systems.
\end{abstract}

\maketitle
\section{Introduction}

The motion of  vortices in two-dimensional incompressible, inviscid flows serves as a canonical example for linking discrete vortex models to continuum fluid dynamics. In this idealization, vorticity is localized at discrete points, reducing the full fluid equations to a finite-dimensional dynamical system. Beyond its classical origins, interest in vortex dipoles, clusters, and collective vortex structures has grown significantly in recent years, driven by developments in quantum fluids, active matter, and hydrodynamic rotor systems. In particular, vortex dipoles and clusters have been directly observed and characterized in trapped Bose-Einstein condensates and two-dimensional quantum fluids~\cite{Neely2010,Freilich2010,white2012,White2014,vsc}, while related decay and collective phenomena arise in superfluid turbulence~\cite{Rooney2011,Goodman2015,Stagg2016}, anomalous vortex fluids~\cite{abanov}, active and soft-matter systems such as microrotors and rotating inclusions~\cite{lushi,yeo,sh1,sh2}, as well as vortex assemblies in the superfluid interiors of neutron stars~\cite{Haskell2015}. These developments provide strong motivation for reduced descriptions of interacting vortex ensembles across a range of physical systems. 
Much of the theoretical and computational literature has focused on unbounded or effectively planar geometries, where translational symmetry and the absence of global image interactions considerably simplify the dynamics~\cite{nc2009}. Compact periodic domains, however, introduce qualitatively new features. In such geometries, each vortex interacts with an infinite lattice of its periodic images, and the resulting motion is governed by the Green’s function of the Laplace operator on the flat torus. This global coupling modifies both few-body motion and collective many-body dynamics, making periodic domains a natural example in which geometry and topology enter explicitly into vortex interactions. Early studies investigated the Hamiltonian structure and stability of periodic vortex arrays, while subsequent work developed lattice-sum representations and analyzed few-vortex dynamics, integrable configurations, and dipolar interactions in doubly periodic domains~\cite{Tkachenko1966,ONeil1989,Dienstfrey2001,WeissMcWilliams1991,Kunin1994,StremlerAref1999,Stremler2010,TsangKanso2013}.
A major analytical advance is the use of the Schottky-Klein prime function, which provides a compact representation of the hydrodynamic Green’s function in multiply connected domains~\cite{Crowdy2005,Crowdy2016,grms}. This framework yields explicit vortex interaction laws incorporating the full periodic geometry and has led to closed-form $N$-vortex equations in terms of special functions~\cite{sakajo2016,sakajo2018,sakajo2019,KrishnamurthySakajo2023}. An equivalent formulation in terms of $q$-special functions has also been developed~\cite{sam3}, enabling efficient simulations and systematic analytic expansions~\cite{sam3,sam5}.

Parallel developments have explored vortex dynamics on curved and compact surfaces from geometric and topological perspectives~\cite{BoattoKoiller2008,GrottaRagazzoGustafssonKoiller2024}. These studies reveal how curvature, topology, and harmonic flows influence vortex motion and stability~\cite{Gustafsson2022,DrivasGlukhovskiyKhesin2024,sam2021,sam2025,sam2026}. For the flat torus, the harmonic component reduces to a constant background flow, allowing a purely interaction-driven description~\cite{GrottaRagazzoGustafssonKoiller2024}. This provides a particularly simple setup for analyzing vortex clusters and binary dynamics in compact periodic domains.

In contrast to the ideal Hamiltonian description, real vortex systems exhibit dissipation due to coupling with additional degrees of freedom. In superfluid helium, this appears as mutual friction arising from quasiparticle scattering off vortex cores~\cite{HallVinen1956a,HallVinen1956b}, with a microscopic description developed by Sonin~\cite{Sonin1987,SoninBook}. Similar effects arise in ultracold Bose-Einstein condensates, where interactions with the thermal cloud lead to vortex decay and spiraling motion~\cite{FedichevShlyapnikov1999,Jackson2008,Anderson2001}. At the level of reduced models, these effects can be captured by augmenting the Hamiltonian vortex velocity with a rotated component, providing a minimal dissipative correction that preserves the geometric structure of the equations while introducing energy decay. Such rotated-velocity formulations are widely used as coarse-grained descriptions of dissipation in vortex dynamics and quantum turbulence~\cite{Sergeev2023}.

Despite extensive work on both periodic vortex dynamics and dissipative vortex motion, a unified analytic treatment of dissipative vortex binaries in compact geometries remains largely unexplored. In this work, we address this problem by combining the exact Hamiltonian formulation on the flat torus with a minimal dissipative extension. The conservative system admits an exact reduction of the two-vortex problem to a single complex relative coordinate, yielding an integrable description. Incorporating dissipation through a rotated-velocity term converts the dynamics into a mixed symplectic-gradient flow with monotonic energy decay for quantized vortices.

We show that, in the local regime, the dissipative vortex binary remains analytically tractable and admits closed-form solutions. Equal same-sign vortices execute outward spiraling motion, while equal opposite-sign pairs (dipoles) undergo finite-time collapse at fixed orientation. For unequal opposite-sign pairs, dissipation induces coupled contraction and rotation, leading to a finite-time nonlinear chirp with $\dot{\omega}\propto\omega^2$, representing the simplest hydrodynamic realization of frequency blow-up driven by dissipative collapse. 

We further derive the leading corrections due to the compact torus geometry by systematically expanding the interaction kernel. These corrections decompose into isotropic and anisotropic contributions, modifying the planar laws and introducing geometry-dependent effects. In particular, the torus breaks the orientation invariance of the dipole and induces a slow angular drift, even in regimes where the planar dynamics would preserve alignment. 

Taken together, these results provide a unified analytic framework for dissipative vortex dynamics on compact domains, capturing the interplay between Hamiltonian structure, dissipation, and global geometry. This framework offers a canonical example for studying vortex relaxation, clustering, and finite-size effects in periodic fluid and superfluid systems.
The remainder of the paper is organized as follows. In Sec.~\ref{cons_summary}, we review the Hamiltonian formulation of point vortices on the flat torus and the exact reduction of the conservative two-vortex problem. In Sec.~\ref{hmdiss}, we introduce the rotated-velocity dissipative correction and show that it produces a mixed symplectic-gradient flow with monotonic Hamiltonian decay. In Sec.~\ref{2vdiss}, we analyze localized dissipative vortex binaries and derive closed-form planar laws for same-sign pairs, equal dipoles, and unequal opposite-sign pairs, including the finite-time chirp law. In Sec.~\ref{tcorr}, we derive the leading torus corrections and show how compact geometry modifies the planar dynamics and  breaks the orientation invariance  of dipole collapse. Finally, the appendices collect explicit perturbative solutions and supporting analytic details.

\section{ Model setup and recap of the conservative dynamics}
\label{cons_summary}
Before turning to dissipative effects, we briefly recap the conservative dynamics of vortex binaries on doubly periodic fluid domains, following Ref.~\cite{KrishnamurthySakajo2023,sam5}. Point-vortex interactions on a doubly periodic inviscid domain, such as the flat torus, admit a Hamiltonian formulation in which interactions are encoded through the Schottky-Klein prime function ~\cite{grms,KrishnamurthySakajo2023} and its $q$-representation~\cite{sam3}. This framework provides an exact description of vortex motion on compact domains and serves as the reference system for the present work. 
Within this formulation, the two-vortex problem reduces to a single complex degree of freedom, yielding explicit expressions for the orbital rotation frequency and dipole translation velocity. Extensions to larger vortex clusters have been explored in Ref.~\cite{sam5}, but here we focus exclusively on the two-vortex dynamics, which forms the foundation for the dissipative analysis developed below.
We consider $N$ point vortices of circulations $\Gamma_j$ in a rectangular torus with periods $2\pi$ and $-\log\rho$ ($0<\rho<1$), and denote their complex positions by
\begin{equation}
w_j = x_j + i y_j,
\qquad
\nu_j = e^{i w_j}.\nn
\end{equation}
The canonical symplectic structure yields the Hamiltonian equations of motion
\begin{equation}
\Gamma_j \dot x_j = \frac{\partial H}{\partial y_j},\nn
\qquad
\Gamma_j \dot y_j = -\frac{\partial H}{\partial x_j},\nn
\end{equation}
or, equivalently, in complex form,
\begin{equation}
\Gamma_j \frac{d w_j}{dt}=-2i\,\frac{\partial H}{\partial \overline{w}_j}.\nn
\end{equation}
In annulus variables, this becomes
\begin{equation}
\Gamma_j \frac{d\overline{w}_j}{dt}=-2\nu_j \frac{\partial H}{\partial \nu_j}.\nn
\end{equation}
The Hamiltonian can be written as
\begin{equation}
H(\nu_1,\dots,\nu_N)
=
-\sum_{1\le j<k\le N}\Gamma_j\Gamma_k\,
G\!\left(\frac{\nu_j}{\nu_k};\sqrt{\rho}\right)
-\frac12\sum_{j=1}^N \Gamma_j^2\,\widehat G(\nu_j;\sqrt{\rho}),
\label{hm}
\end{equation}
where the pair hydrodynamic Green function is
\begin{equation}
G(\zeta;\sqrt{\rho})
=
\frac{1}{2\pi}\log\!\bigl|P(\zeta,\sqrt{\rho})\bigr|
-\frac{1}{4\pi}\log|\zeta|
+\frac{1}{4\pi\log\rho}\bigl(\log|\zeta|\bigr)^2.
\end{equation}
For the flat torus, the regular part of the Green's function $\widehat G$ reduces to a geometry-dependent constant~\cite{GrottaRagazzoGustafssonKoiller2024,Gustafsson2022}. so it does not contribute to the equations of motion. The resulting $N$-vortex evolution is therefore governed entirely by pair interactions:
\begin{equation}
\frac{d\overline{w}_j}{dt}
=
\frac{1}{2\pi}
\sum_{\substack{k=1\\k\neq j}}^{N}
\Gamma_k\,K\!\left(\frac{\nu_j}{\nu_k},\sqrt{\rho}\right)
-\frac{1}{4\pi}
\sum_{\substack{k=1\\k\neq j}}^{N}\Gamma_k
+\frac{1}{2\pi\log\rho}
\sum_{\substack{k=1\\k\neq j}}^{N}
\Gamma_k\log\!\left|\frac{\nu_j}{\nu_k}\right|,
\label{eq:summary_dyneq}
\end{equation}
with $\log|\nu_j/\nu_k|=-(y_j-y_k)$. A convenient closed-form expression for the interaction kernel is given by~\cite{sam3}:
\begin{equation}
K(\zeta,\sqrt{\rho})
=
\frac{1}{1-\zeta}
+\frac{1}{\log\rho}
\left[
\psi_\rho\!\left(\frac{\log(1/\zeta)}{\log\rho}\right)
-
\psi_\rho\!\left(\frac{\log\zeta}{\log\rho}\right)
\right],
\label{eq:summary_kernel}
\end{equation}
where $\psi_\rho(z)=\frac{d}{dz}\log\Gamma_\rho(z)$ is the $q$-digamma function associated with the $q$-gamma function
\begin{equation}
\Gamma_\rho(z)
=
(1-\rho)^{1-z}
\prod_{n=0}^\infty
\frac{1-\rho^{n+1}}{1-\rho^{n+z}}.
\end{equation}
Equation~\eqref{eq:summary_dyneq} makes explicit the three ingredients of the torus dynamics: the singular pair interaction, the constant background contribution, and the geometry-dependent logarithmic correction. It is convenient to write the dynamics as
\begin{equation}
\frac{d\overline{w}_j}{dt}
=
\sum_{\substack{k=1\\k\neq j}}^N
\Gamma_k\,F\!\left(\frac{\nu_j}{\nu_k}\right),
\qquad
F(\zeta)
=
\frac{1}{2\pi}K(\zeta,\sqrt{\rho})
-\frac{1}{4\pi}
+\frac{1}{2\pi\log\rho}\log|\zeta|.
\label{eq:summary_F}
\end{equation}
Using
\begin{equation}
K(1/\zeta)=1-K(\zeta),
\qquad
\log|1/\zeta|=-\log|\zeta|,\nn
\end{equation}
one obtains the crucial antisymmetric relation
\begin{equation}
F(1/\zeta)=-F(\zeta),
\label{eq:summary_antisym}
\end{equation}
which is the key structural identity of the nondissipative problem. It implies conservation of the Hamiltonian and of the circulation-weighted centroid
\begin{equation}
C=\sum_{j=1}^N \Gamma_j w_j,
\qquad
\overline C=\sum_{j=1}^N \Gamma_j \overline{w}_j,
\end{equation}
equivalently the translational invariants $\sum_j \Gamma_j x_j$ and $\sum_j \Gamma_j y_j$. The total circulation $\sum_j\Gamma_j$  is trivially  conserved. By contrast with the planar problem, the flat torus does not possess a continuous rotational symmetry, so there is no additional angular-momentum-type invariant.

The same antisymmetric feature yields a particularly simple description of binary motion. For the relative coordinate $w_{12}=w_1-w_2$ one finds
\begin{equation}
\frac{d\overline{w}_{12}}{dt}
=
(\Gamma_1+\Gamma_2)\,
F\!\left(\frac{\nu_1}{\nu_2}\right),
\label{eq:summary_binary_w12}
\end{equation}
so the two-vortex problem closes exactly. Introducing
\begin{equation}
\eta=\frac{\nu_1}{\nu_2}=e^{i(w_1-w_2)},
\end{equation}
the relative dynamics reduces to the scalar complex equation
\begin{equation}
\dot\eta=i\,\eta\,\Gamma_{\rm tot}\,F(\eta),
\qquad
\Gamma_{\rm tot}=\Gamma_1+\Gamma_2,
\label{eq:summary_eta}
\end{equation}
with implicit solution
\begin{equation}
\int_{\eta_0}^{\eta(t)} \frac{d\zeta}{i\zeta F(\zeta)}
=
\Gamma_{\rm tot}(t-t_0).
\label{eq:summary_quadrature}
\end{equation}
Thus the nondissipative binary on the flat torus is completely integrable. In the dipole case $\Gamma_{\rm tot}=0$, Eq.~\eqref{eq:summary_binary_w12} shows immediately that
\begin{equation}
\dot w_{12}=0,
\end{equation}
so the separation is frozen and the pair translates rigidly with constant relative configuration. For $\Gamma_{\rm tot}\neq 0$, the pair executes nontrivial relative motion, and stationary separations are determined by the roots
\begin{equation}
F(\eta_\ast)=0.
\end{equation}
Writing $\eta=r e^{i\theta}$ and $K=K_R+iK_I$, Eq.~\eqref{eq:summary_eta} gives
\begin{equation}
\frac{\dot r}{r}
=
-\Gamma_{\rm tot}\,\Im F(\eta)
=
-\frac{\Gamma_{\rm tot}}{2\pi}K_I(\eta,\sqrt{\rho}),
\label{eq:summary_r}
\end{equation}
and
\begin{equation}
\dot\theta
=
\Gamma_{\rm tot}\,\Re F(\eta)
=
\Gamma_{\rm tot}
\left[
\frac{1}{2\pi}K_R(\eta,\sqrt{\rho})
-\frac{1}{4\pi}
+\frac{\log r}{2\pi\log\rho}
\right].
\label{eq:summary_theta}
\end{equation}
These equations cleanly separate the two effects of the torus kernel: the imaginary part $K_I$ controls radial drift in the annulus variable, while the real part $K_R$ controls phase rotation. Constant-$r$ orbits satisfy $K_I=0$, in which case the annulus angular velocity (this is different from the usual Euclidean angular velocity) is
\begin{equation}
\Omega_\eta
=
\Gamma_{\rm tot}
\left[
\frac{1}{2\pi}K_R(\eta,\sqrt{\rho})
-\frac{1}{4\pi}
+\frac{\log r}{2\pi\log\rho}
\right].
\end{equation}
For equal like-signed vortices, $\Gamma_1=\Gamma_2=\gamma$, the generic equations reduce to
\begin{equation}
\frac{\dot r}{r}
=
-\frac{\gamma}{\pi}K_I(\eta,\sqrt{\rho}),
\qquad
\Omega_\eta
=
\frac{\gamma}{\pi}
\left[
K_R(\eta,\sqrt{\rho})-\frac12+\frac{\log r}{\log\rho}
\right].
\end{equation}
Because $\theta$ is the natural phase of the annulus variable rather than the physical polar angle in the $(x,y)$ plane, it is useful to also define the Euclidean orbital frequency from the relative vector $\Delta=w_1-w_2=x_{12}+iy_{12}$:
\begin{equation}
\Omega_E
=
\frac{x_{12}\dot y_{12}-y_{12}\dot x_{12}}{x_{12}^2+y_{12}^2}.
\end{equation}
Using Eq.~\eqref{eq:summary_F}, one obtains the exact expression
\begin{equation}
\Omega_E
=
-\frac{\Gamma_{\rm tot}}{2\pi}
\frac{
x_{12}K_I(\eta,\sqrt{\rho})
+y_{12}K_R(\eta,\sqrt{\rho})
-\tfrac12 y_{12}
-\dfrac{y_{12}^2}{\log\rho}
}{x_{12}^2+y_{12}^2}.
\label{eq:summary_OmegaE}
\end{equation}
Equation~\eqref{eq:summary_OmegaE} shows that the observed orbital frequency depends not only on the interaction kernel but also on the instantaneous geometry of the relative separation. In particular, it differs from the annulus frequency $\dot\theta$ and receives distinct contributions from $x_{12}K_I$, $y_{12}K_R$, and the explicit torus correction proportional to $y_{12}^2/\log\rho$. Numerically, this expression accurately reproduces the instantaneous orbital frequency of representative equal-vortex trajectories, with residuals at the $10^{-7}$ level, as demonstrated in Ref.~\cite{sam5}.\\
Finally, in the dipole case $\Gamma_1=\gamma$, $\Gamma_2=-\gamma$, the two vortices move with a common constant relative displacement and a translational velocity
\begin{equation}
\frac{d\overline{w}_1}{dt}
=
\frac{d\overline{w}_2}{dt}
=
-\gamma\,F(\eta),
\end{equation}
or, in real form,
\begin{equation}
\dot x
=
-\frac{\gamma}{4\pi}
\left[
2K_R(\eta,\sqrt{\rho})-1-\frac{2y_{12}}{\log\rho}
\right],
\qquad
\dot y
=
\frac{\gamma}{2\pi}K_I(\eta,\sqrt{\rho}).
\label{eq:summary_dipole_velocity}
\end{equation}
Hence, unlike the planar dipole, the propagation speed on the flat torus is not determined solely by the separation magnitude, but depends explicitly on the global periodic geometry through both the kernel and the logarithmic correction. This Hamiltonian conservative problem provides the natural  starting point for studying the dissipative dynamics developed in the following sections.
\section{Dissipative flow equations}
\label{hmdiss}
\begin{figure}[t]
\centering
\setlength{\abovecaptionskip}{2pt}
\setlength{\belowcaptionskip}{0pt}

\begin{subfigure}{0.46\textwidth}
\centering
\includegraphics[height=0.17\textheight]{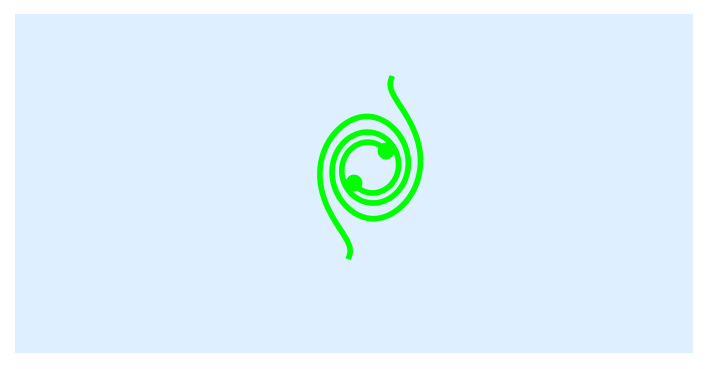}
\caption{Same-sign.}
\end{subfigure}
\hfill
\begin{subfigure}{0.46\textwidth}
\centering
\includegraphics[height=0.17\textheight]{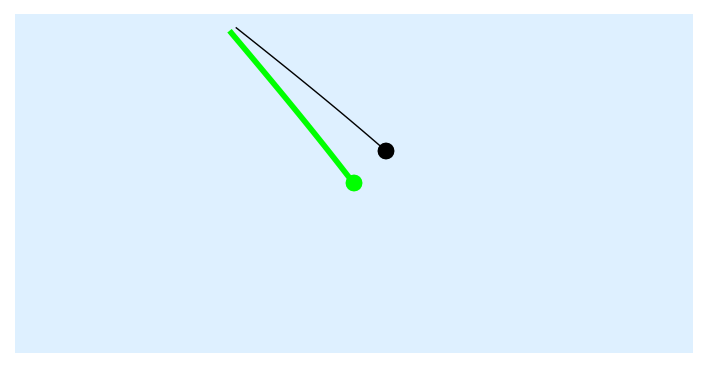}
\caption{Dipole.}
\end{subfigure}

\vspace{0.05cm}

\begin{subfigure}{0.46\textwidth}
\centering
\includegraphics[height=0.17\textheight]{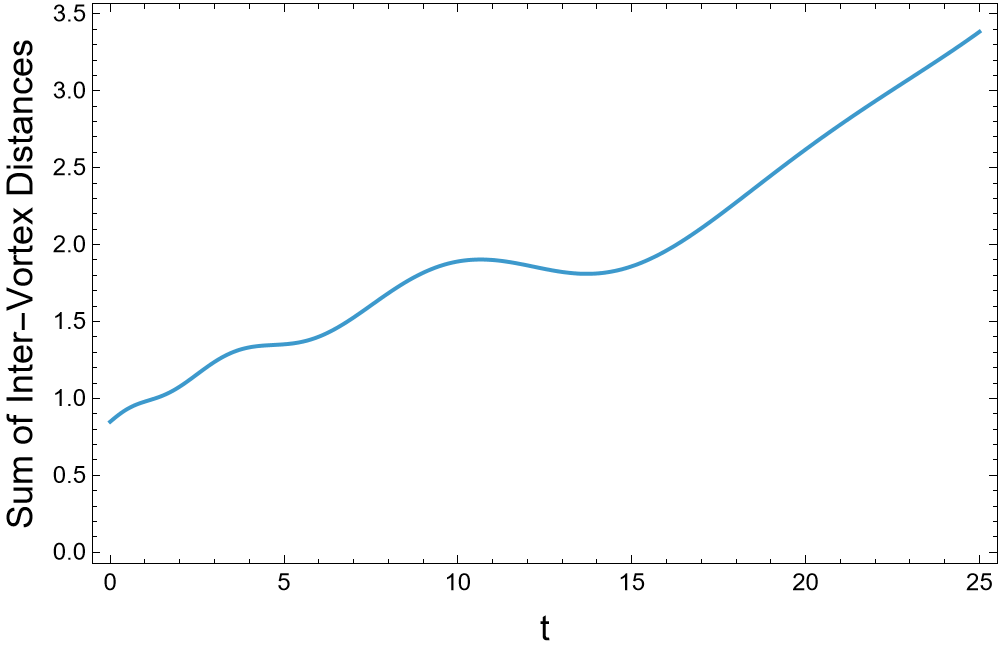}
\caption{Distance (same sign).}
\end{subfigure}
\hfill
\begin{subfigure}{0.46\textwidth}
\centering
\includegraphics[height=0.17\textheight]{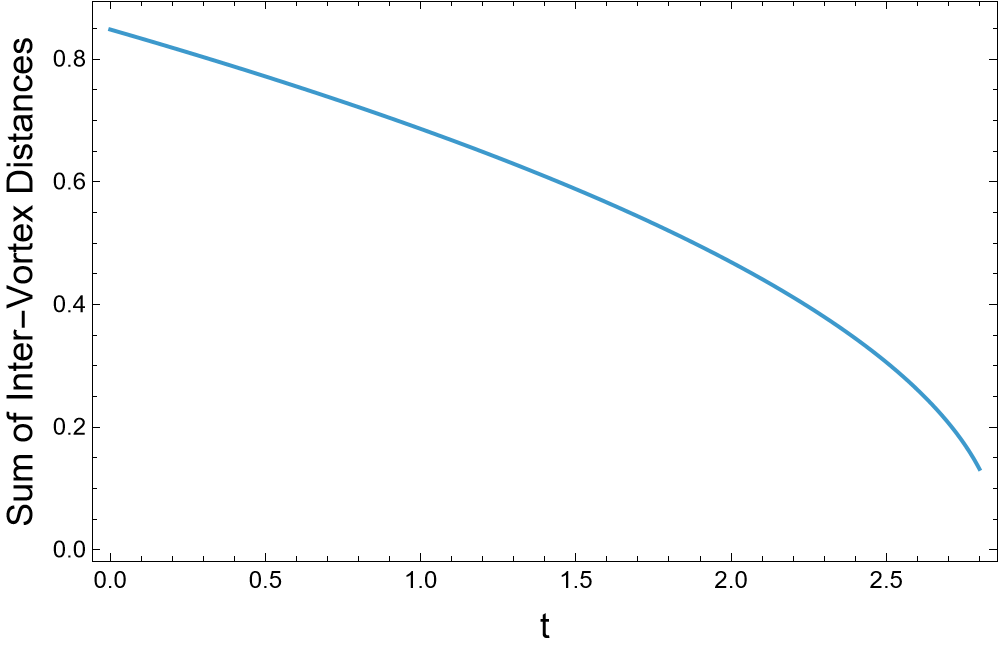}
\caption{Distance (dipole).}
\end{subfigure}

\vspace{0.05cm}

\begin{subfigure}{0.46\textwidth}
\centering
\includegraphics[height=0.17\textheight]{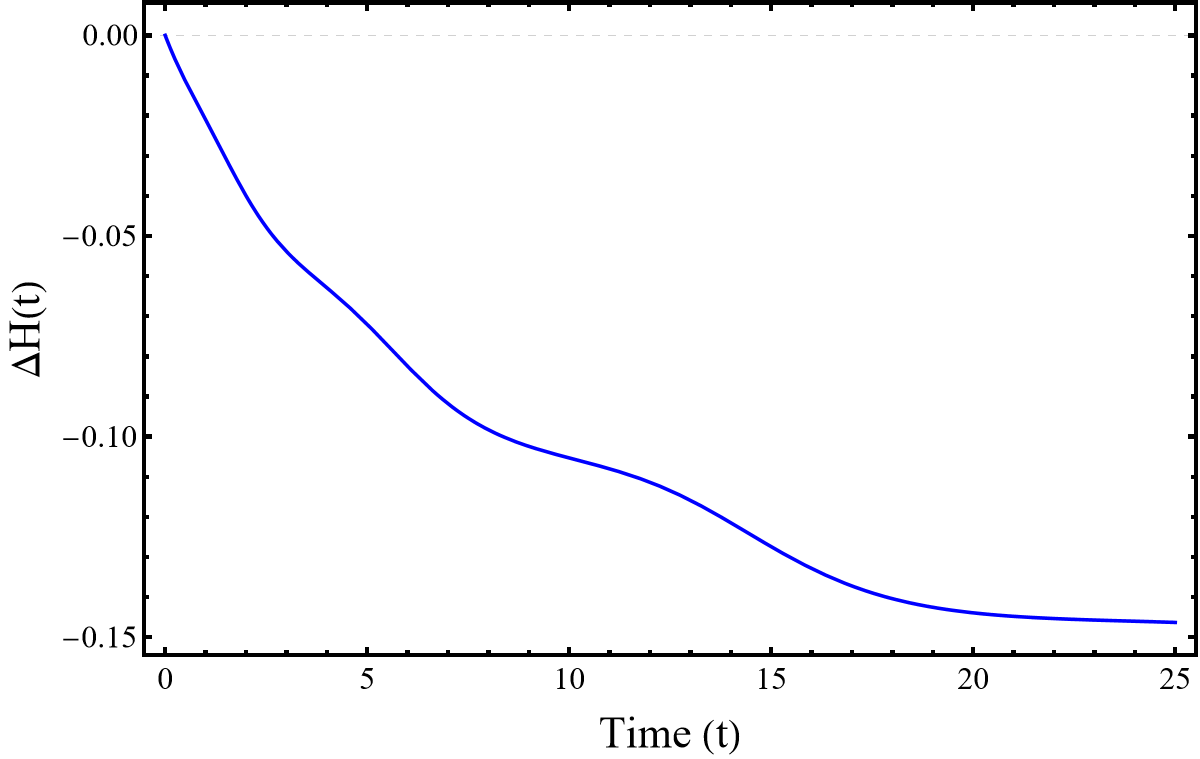}
\caption{$\Delta H$ (same sign).}
\end{subfigure}
\hfill
\begin{subfigure}{0.46\textwidth}
\centering
\includegraphics[height=0.17\textheight]{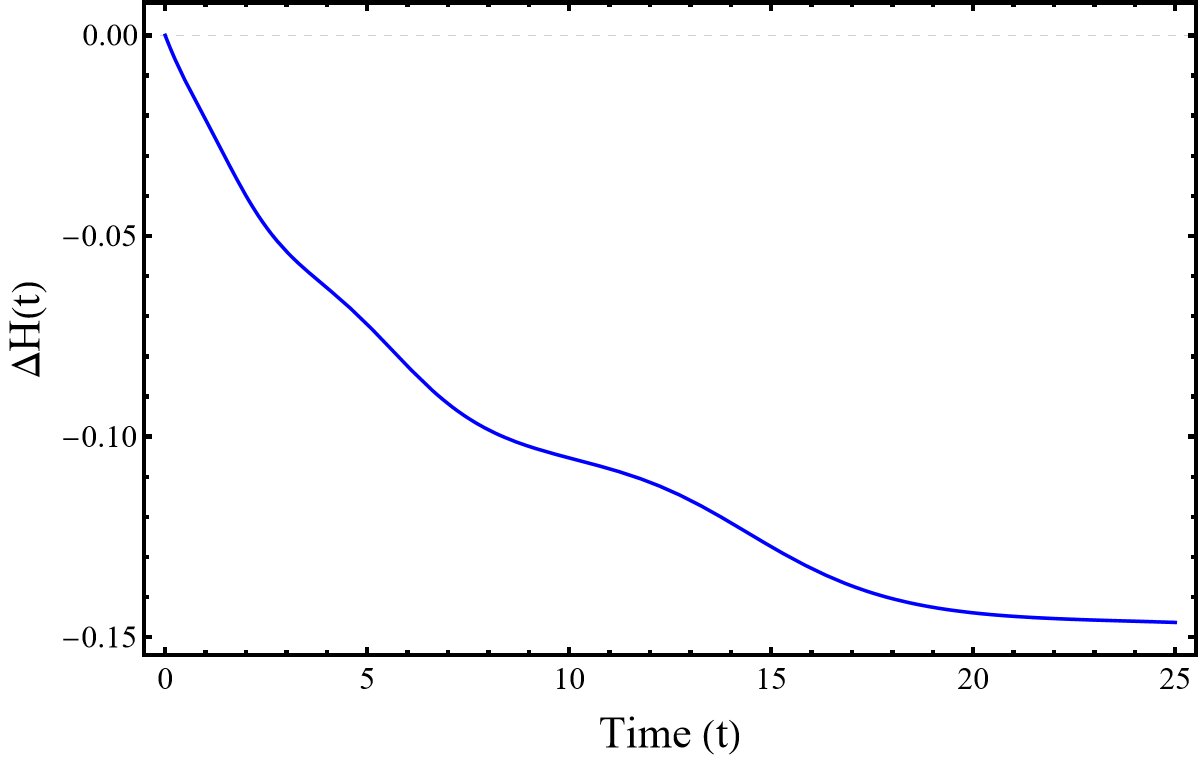}
\caption{$\Delta H$ (dipole).}
\end{subfigure}

\caption{
Dissipative two-vortex dynamics on a rectangular flat torus 
($\rho=e^{-\pi}$, $\gamma=0.1$) with initial positions 
$(\pi,\pi/2)$ and $(\pi+0.3,\pi/2+0.3)$.
Left: same-sign vortices; right: opposite-sign (dipole).
Top: trajectories; middle: pair separation; bottom: Hamiltonian decay.
}
\label{fig:2v_diss_compare}
\end{figure}
\begin{figure}[t]
\centering
\setlength{\abovecaptionskip}{3pt}
\setlength{\belowcaptionskip}{0pt}

\begin{subfigure}{0.47\textwidth}
\centering
\includegraphics[width=\linewidth]{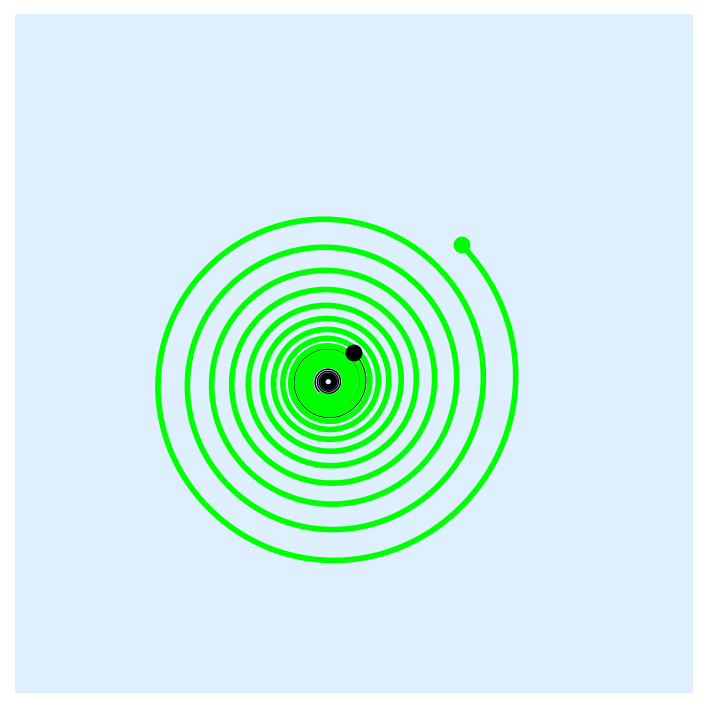}
\caption{Trajectories.}
\end{subfigure}
\hfill
\begin{subfigure}{0.47\textwidth}
\centering
\includegraphics[width=\linewidth]{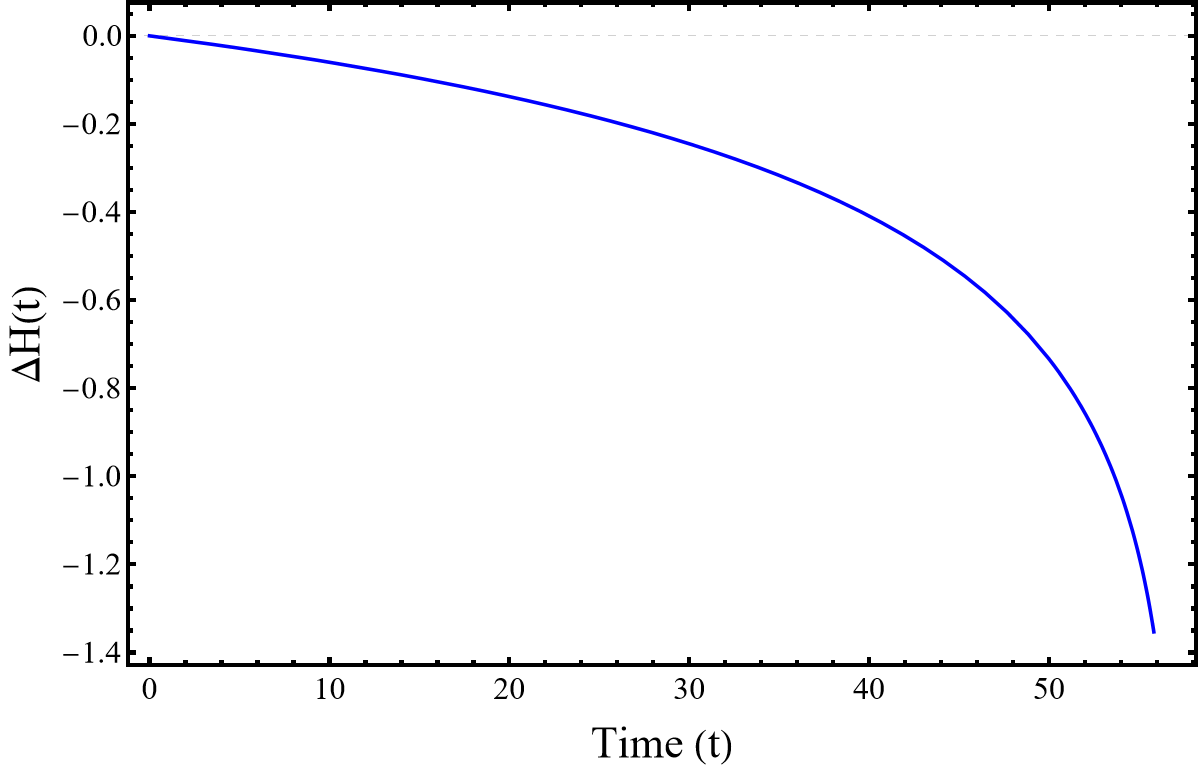}
\caption{$\Delta H(t)$.}
\end{subfigure}

\vspace{0.12cm}

\begin{subfigure}{0.47\textwidth}
\centering
\includegraphics[width=\linewidth]{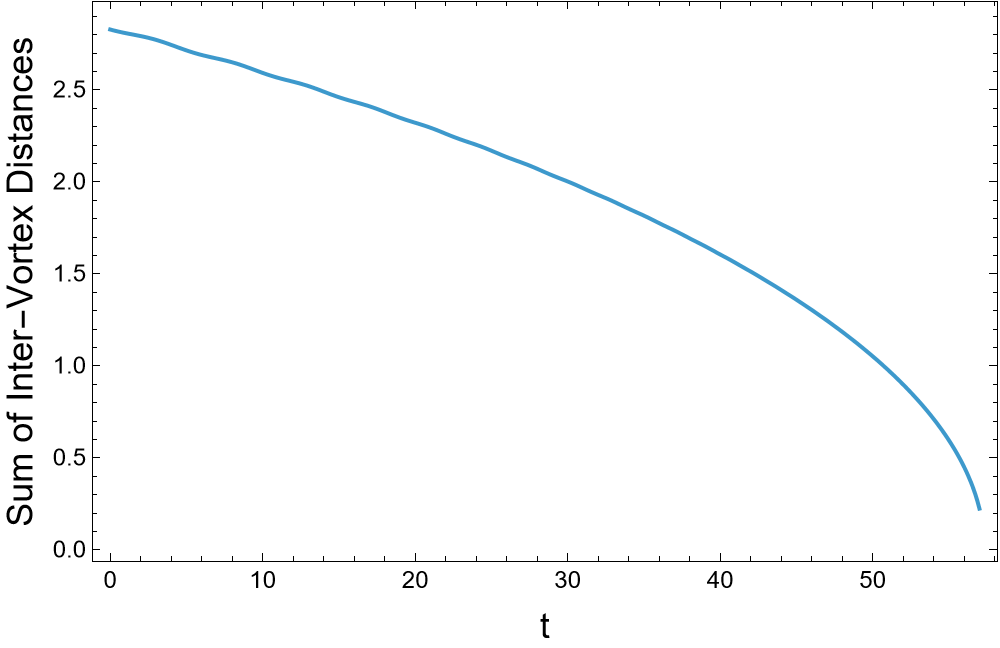}
\caption{Pair-distance measure.}
\end{subfigure}
\hfill
\begin{subfigure}{0.47\textwidth}
\centering
\includegraphics[width=\linewidth]{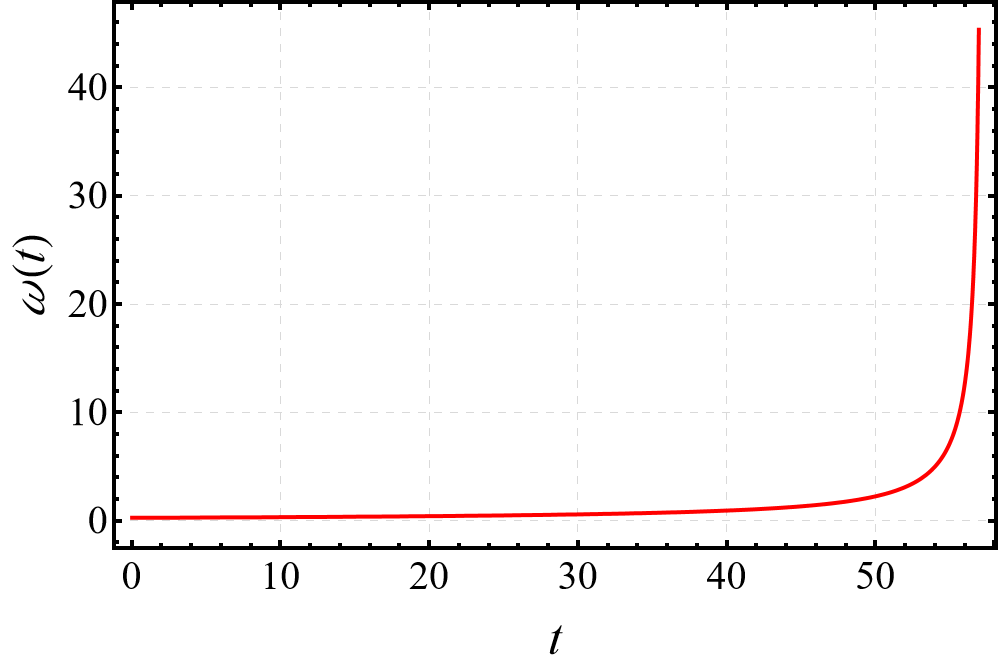}
\caption{$\omega(t)$.}
\end{subfigure}

\vspace{0.12cm}

\begin{subfigure}{0.47\textwidth}
\centering
\includegraphics[width=\linewidth]{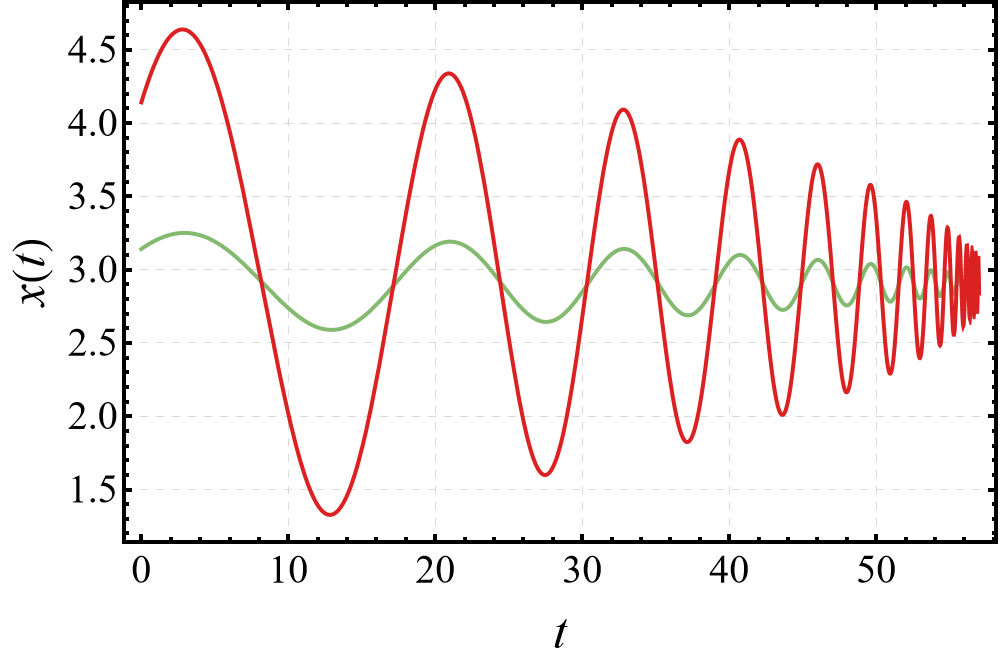}
\caption{$x(t)$.}
\end{subfigure}
\hfill
\begin{subfigure}{0.47\textwidth}
\centering
\includegraphics[width=\linewidth]{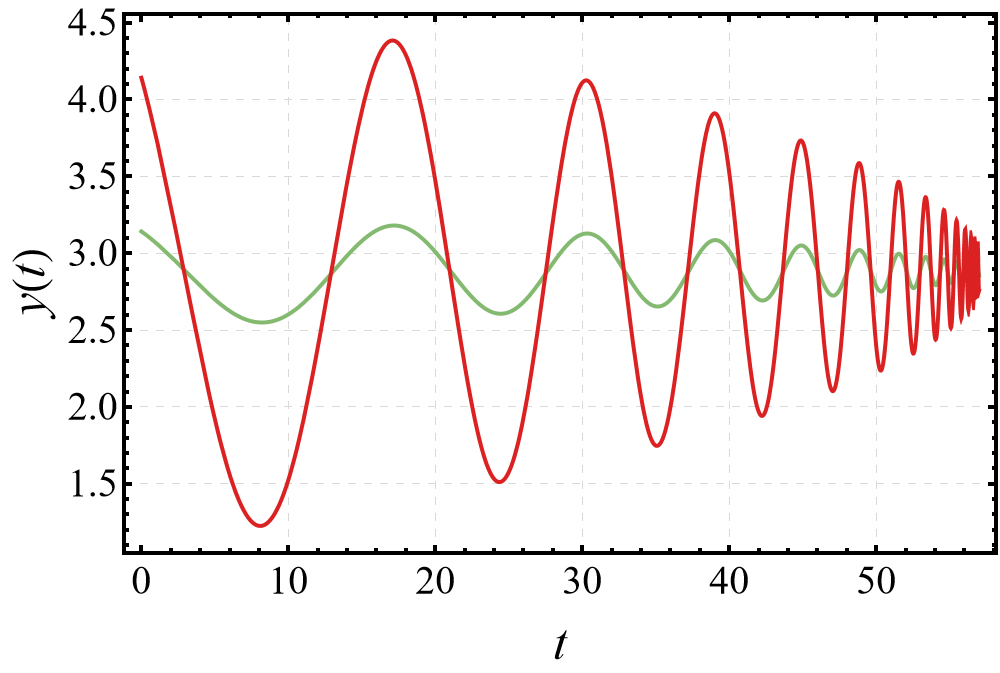}
\caption{$y(t)$.}
\end{subfigure}

\caption{
Dissipative two-vortex dynamics on the flat square torus ($\rho = e^{-2\pi}$, $\gamma=0.02$).
Top: trajectories showing inward spiraling motion and monotonic decay of the Hamiltonian $\Delta H(t)$.
Middle: contraction of the pair separation, quantified by the sum of inter-vortex distances, together with the growth of the instantaneous angular frequency $\omega(t)$.
Bottom: coordinate dynamics $(x(t),y(t))$, exhibiting oscillatory motion with increasing frequency as the vortices approach a lower-energy configuration.
}
\label{figsquare}
\end{figure}
Dissipative corrections to vortex motion arise from the interaction of vortices with additional degrees of freedom in superfluids and finite-temperature quantum fluids. In contrast to ideal Hamiltonian motion governed by the Magnus force, real systems exhibit \emph{mutual friction}, whereby scattering of thermal excitations (phonons and rotons) off vortex cores transfers momentum and generates drag. This mechanism, introduced phenomenologically by Gorter and Mellink and quantified in the experiments of Hall and Vinen, leads to a modified force balance in which both longitudinal and transverse friction coefficients supplement the Magnus dynamics. Microscopic analyses based on quasiparticle scattering show that these dissipative effects can be represented, at a coarse-grained level, by additional velocity components both parallel and transverse to the Hamiltonian flow.

Analogous behavior arises in ultracold Bose-Einstein condensates, where coupling between vortices and the thermal cloud induces energy dissipation and vortex decay. At the level of reduced models, these effects are captured by a minimal extension of point-vortex dynamics,
\begin{equation}
\dot{\mathbf r}_i
=
\mathbf v_i^{(0)}
-
\alpha s_i\,\hat{\mathbf z}\times \mathbf v_i^{(0)},
\end{equation}
which provides a coarse-grained representation of the leading transverse drift induced by thermal excitations. More generally, mutual friction is a generic mechanism underlying dissipation in bosonic superfluids and quantum turbulence.
We introduce dissipation directly at the level of the Hamiltonian point-vortex dynamics. In complex variables, the conservative equations are
\begin{equation}
\Gamma_j \dot{w}_j=-2i\,\frac{\partial H}{\partial \overline{w}_j},
\qquad
\Gamma_j \dot{\overline{w}}_j=2i\,\frac{\partial H}{\partial w_j},
\label{eq:diss_ham}
\end{equation}
which define a purely symplectic flow and conserve the Hamiltonian $H$. To model friction, we add to each vortex velocity a component rotated by $90^\circ$,
\begin{equation}
\mathbf{v}_j^{\rm diss}
=
\mathbf{v}_j-\gamma \kappa_j\,\hat{\mathbf z}\times \mathbf{v}_j,
\label{eq:diss_real}
\end{equation}
where $\gamma>0$ is a dimensionless dissipation coefficient and $\kappa_j=\pm1$ denotes the vortex circulation. In complex form, the rotation $\hat{\mathbf z}\times \mathbf{v}_j=(-v_{y,j},v_{x,j})$ is represented by multiplication by $i$, so that
\begin{equation}
\dot{w}_j^{\rm diss}=(1-i\gamma\kappa_j)\dot{w}_j,
\qquad
\dot{\overline{w}}_j^{\rm diss}=(1+i\gamma\kappa_j)\dot{\overline{w}}_j.
\label{eq:diss_complex}
\end{equation}
Substituting Eq.~\eqref{eq:diss_ham}, we obtain
\begin{equation}
\dot{w}_j^{\rm diss}
=
-\frac{2i}{\Gamma_j}\frac{\partial H}{\partial \overline{w}_j}
-\frac{2\gamma\kappa_j}{\Gamma_j}\frac{\partial H}{\partial \overline{w}_j},
\qquad
\dot{\overline{w}}_j^{\rm diss}
=
\frac{2i}{\Gamma_j}\frac{\partial H}{\partial w_j}
-\frac{2\gamma\kappa_j}{\Gamma_j}\frac{\partial H}{\partial w_j}.
\label{eq:diss_grad}
\end{equation}
The dissipative evolution therefore splits naturally into a Hamiltonian part and a gradient part: the first term generates motion tangent to the level sets of $H$, while the second drives motion down the energy landscape. For a real-valued Hamiltonian $H(w,\overline{w})$, its rate of change along the dissipative flow is
\begin{equation}
\frac{dH}{dt}
=
\sum_j
\left(
\frac{\partial H}{\partial w_j}\dot{w}_j^{\rm diss}
+
\frac{\partial H}{\partial \overline{w}_j}\dot{\overline{w}}_j^{\rm diss}
\right).
\end{equation}
The symplectic contributions cancel identically, leaving
\begin{equation}
\frac{dH}{dt}
=
-4\gamma
\sum_j
\frac{\kappa_j}{\Gamma_j}
\frac{\partial H}{\partial w_j}
\frac{\partial H}{\partial \overline{w}_j}.
\end{equation}
Since $H$ is real,
\[
\frac{\partial H}{\partial \overline{w}_j}
=
\overline{\frac{\partial H}{\partial w_j}},
\]
and therefore
\begin{equation}
\frac{dH}{dt}
=
-4\gamma
\sum_j
\frac{\kappa_j}{\Gamma_j}
\left|
\frac{\partial H}{\partial w_j}
\right|^2.
\label{eq:dHdt_general}
\end{equation}
In particular, for quantized vortices with unit circulation magnitude, so that $\Gamma_j=\kappa_j=\pm1$, this reduces to
\begin{equation}
\frac{dH}{dt}
=
-4\gamma
\sum_j
\left|
\frac{\partial H}{\partial w_j}
\right|^2
\le 0.
\label{eq:dHdt_equal}
\end{equation}
Thus the dissipative correction converts the Hamiltonian vortex dynamics into a mixed symplectic-gradient flow with monotonic energy decay for $\gamma>0$. More generally, monotonicity requires $\kappa_j/\Gamma_j>0$, which is naturally satisfied when $\kappa_j$ is identified with the sign of the circulation. For unequal-strength point vortices, however, the precise mobility factor multiplying the gradient term is not fixed uniquely by Hamiltonian structure alone, but depends on the underlying microscopic drag mechanism. In that sense, the choices $c_j=\kappa_j$ and $c_j=\Gamma_j$ should be regarded as distinct phenomenological choices rather than universal extensions \cite{Haskell2015}.
\section{Dissipative dynamics of the vortex binary in doubly periodic fluid domain}
\label{2vdiss}
As formulated in the last section, we incorporate dissipation by augmenting the Hamiltonian point-vortex velocity with a rotated drift. Starting from the conservative evolution
\begin{equation}
\frac{d\overline{w}_j}{dt}=F_j(\{w\}),
\end{equation}
the dissipative dynamics is taken to be
\begin{equation}
\frac{d\overline{w}_j^{\rm diss}}{dt}
=
(1+i\gamma \kappa_j)\frac{d\overline{w}_j}{dt},
\label{eq:binary_diss_rule}
\end{equation}
where $\gamma>0$ is the dissipation strength and $\kappa_j=\operatorname{sgn}(\Gamma_j)$. For a binary, the antisymmetry $F(1/\zeta)=-F(\zeta)$ implies an exact closed equation for the relative coordinate
\begin{equation}
w=w_1-w_2,
\qquad
\zeta=\frac{\nu_1}{\nu_2}=e^{iw}.\nn
\end{equation}
In the local regime $|w|\ll1$, the interaction kernel has the expansion (Appendix \ref{app2})
\begin{equation}
F(w,\bar w)
=
-\frac{i}{2\pi w}
+
A(\rho)\,w
+
B(\rho)\,\bar w
+
\cdots,
\label{eq:binary_local_kernel}
\end{equation}
with
\begin{equation}
A(\rho)
=
\frac{i}{24\pi \log^2\rho}
\left[
\log\rho\,(6+\log\rho)-24\,\psi_\rho^{(1)}(1)
\right],
\qquad
B(\rho)=-\frac{i}{4\pi\log\rho}.
\end{equation}
To leading order, the binary dynamics is governed entirely by the singular term, which will be the focus of this section. Corrections arising from the toroidal geometry are addressed in subsequent sections.
For two equal like-signed vortices, $\Gamma_1=\Gamma_2=\Gamma>0$, the relative (conservative) equation reduces to
\begin{equation}
\frac{d\overline w}{dt}
=
-\frac{i\Gamma}{\pi w}.
\end{equation}
Writing $w=r e^{i\theta}$, the conservative motion satisfies
\begin{equation}
\dot r=0,
\qquad
\dot\theta=\frac{\Gamma}{\pi r^2},
\end{equation}
so the pair rotates rigidly at fixed separation. With dissipation, Eq.~\eqref{eq:binary_diss_rule} gives
\begin{equation}
\dot r=\frac{\Gamma\gamma}{\pi r},
\qquad
\dot\theta=\frac{\Gamma}{\pi r^2},
\end{equation}
and hence
\begin{equation}
r^2(t)=r_0^2+\frac{2\Gamma\gamma}{\pi}t,
\qquad
\theta(t)=\theta_0+\frac{1}{2\gamma}
\log\!\left(1+\frac{2\Gamma\gamma}{\pi r_0^2}t\right).
\label{eq:samesign_spiral}
\end{equation}
Thus the dissipative same-sign pair executes an outward spiral, with $r(t)\sim t^{1/2}$ and $\dot\theta\sim t^{-1}$ at late times.\\ For an equal circulation dipole, $\Gamma_1=\Gamma$ and $\Gamma_2=-\Gamma$, the conservative velocities are identical, so the relative coordinate is frozen (in the conservative case),
\begin{equation}
\frac{d\overline w}{dt}=0,
\end{equation}
and the pair translates rigidly. The dissipative correction yields
\begin{equation}
\frac{d\overline w}{dt}^{\rm diss}
=
-\frac{\Gamma\gamma}{\pi w},
\end{equation}
so that
\begin{equation}
\dot r=-\frac{\Gamma\gamma}{\pi r},
\qquad
\dot\theta=0,
\end{equation}
and therefore
\begin{equation}
r^2(t)=r_0^2-\frac{2\Gamma\gamma}{\pi}t,
\qquad
\theta(t)=\theta_0.
\label{eq:dipole_collapse}
\end{equation}
The dipole thus retains its orientation and translational character, while the separation decreases monotonically.\\ For an opposite-sign unequal-strength pair, $\Gamma_1=\Gamma_a>0$ and $\Gamma_2=-\Gamma_b<0$, the conservative relative dynamics is
\begin{equation}
\frac{d\overline w}{dt}
=
-\frac{i(\Gamma_a-\Gamma_b)}{2\pi w},
\end{equation}
which implies
\begin{equation}
\dot r=0,
\qquad
\dot\theta=\frac{\Gamma_a-\Gamma_b}{2\pi r^2}.
\end{equation}
Thus, unlike the equal dipole, the unequal pair rotates rigidly at fixed separation. Including dissipation gives
\begin{equation}
\frac{d\overline w}{dt}^{\rm diss}
=
\frac{-\gamma(\Gamma_a+\Gamma_b)-i(\Gamma_a-\Gamma_b)}{2\pi w},
\end{equation}
or equivalently
\begin{equation}
\dot r
=
-\frac{\gamma(\Gamma_a+\Gamma_b)}{2\pi r},
\qquad
\dot\theta
=
\frac{\Gamma_a-\Gamma_b}{2\pi r^2}.
\label{eq:unequal_pair_local}
\end{equation}
Hence the pair undergoes simultaneous contraction and rotation. Integrating,
\begin{equation}
r^2(t)=r_0^2-\frac{\gamma(\Gamma_a+\Gamma_b)}{\pi}t,
\end{equation}
and
\begin{equation}
\theta(t)
=
\theta_0
-
\frac{\Gamma_a-\Gamma_b}{2\gamma(\Gamma_a+\Gamma_b)}
\log\!\left(
1-\frac{\gamma(\Gamma_a+\Gamma_b)}{\pi r_0^2}t
\right).
\label{eq:unequal_pair_theta}
\end{equation}
The equal-dipole limit is recovered when $\Gamma_a=\Gamma_b$, while the conservative rigid binary is recovered for $\gamma=0$. To derive an equation for the rate of change of frequency, it is convenient to introduce two notations
\begin{equation}
\Sigma=\Gamma_a+\Gamma_b,
\qquad
\Delta=\Gamma_a-\Gamma_b,
\end{equation}
for which Eq.~\eqref{eq:unequal_pair_local} becomes
\begin{equation}
\dot r=-\frac{\gamma\Sigma}{2\pi r},
\qquad
\omega(t)\equiv \dot\theta=\frac{\Delta}{2\pi r^2}.
\end{equation}
The radial equation gives
\begin{equation}
r^2(t)=r_0^2-\frac{\gamma\Sigma}{\pi}t,
\qquad
t_*=\frac{\pi r_0^2}{\gamma\Sigma},
\end{equation}
and therefore
\begin{equation}
\omega(t)
=
\frac{\Delta}{2\pi\left(r_0^2-\frac{\gamma\Sigma}{\pi}t\right)}
=
\frac{\omega_0}{1-t/t_*},
\qquad
\omega_0=\frac{\Delta}{2\pi r_0^2}.
\label{eq:chirp_explicit}
\end{equation}
Differentiating yields the closed chirp equation
\begin{equation}
\dot\omega
=
\frac{2\gamma\Sigma}{\Delta}\,\omega^2.
\label{eq:chirp_law}
\end{equation}
Thus the dissipative unequal opposite-sign binary exhibits a finite-time nonlinear chirp, with frequency blow-up driven by radial collapse. Numerical results are in agreement with the local laws \eqref{eq:unequal_pair_local}-\eqref{eq:chirp_law}; the small oscillatory deviations observed in the full torus dynamics arise from the higher-order terms in Eq.~\eqref{eq:binary_local_kernel}, which encode anisotropic and nonlocal corrections due to the periodic geometry. The vortex system exhibits a chirp law of the form
\(
\dot\omega \propto \omega^{2},
\)
representing the simplest nonlinear frequency blow-up arising from dissipative collapse. More generally, inspiral dynamics across physical systems display finite-time frequency divergence with scaling exponents set by the dominant radiation mechanism. In gravitational-wave-driven compact binaries, energy loss is quadrupolar, leading to the well-known scaling
\(
\dot\omega \propto \omega^{11/3},
\)
derived in the post-Newtonian framework~\cite{Peters1964}. By contrast, electromagnetic dipole-dominated inspirals exhibit
\(
\dot\omega \propto \omega^{3},
\)
as follows from leading-order radiation reaction in charged systems~\cite{Verma2025}. The vortex chirp thus represents the lowest-order member of this hierarchy, arising from hydrodynamic interactions and local dissipation rather than radiation to infinity.
\section{Geometric corrections to dissipative binary dynamics}
\label{tcorr}
We now compute the leading geometric corrections to the local dissipative binary dynamics. For a localized pair, $|w|\ll 1$, the doubly periodic interaction kernel admits the expansion (see Ref.~\cite{sam5} and Appendix~\ref{app3} for details)
\begin{equation}
F(w,\overline w)
=
-\frac{i}{2\pi w}
+i a(\rho) w
+i b(\rho)\overline w
+\cdots ,
\label{eq:torus_local_expansion}
\end{equation}
where $a(\rho)$ and $b(\rho)$ are real coefficients fixed by the torus modulus,
\begin{equation}
a(\rho)
=
\frac{1}{24\pi \log^2\rho}
\left[
\log\rho\,(6+\log\rho)
-
24\,\psi_\rho^{(1)}(1)
\right],
\qquad
b(\rho)
=
-\frac{1}{4\pi\log\rho}.\nn
\end{equation} 
where $\psi_\rho^{(1)}(1)$ denotes the first derivative of the $q$-digamma function $\psi_\rho(z)=\frac{d}{dz}\log\Gamma_\rho(z)$ evaluated at $z=1$, with $\Gamma_\rho(z)$ the $q$-gamma function associated with the torus modulus $\rho$. The singular term in Eq.~(\ref{eq:torus_local_expansion}) is the universal planar vortex interaction, while the terms proportional to $a$ and $b$ give the first geometry-dependent corrections.  It will be helpful in the following analysis to use polar variables
\begin{equation}
w=r e^{i\theta},\qquad R=r^2,
\label{newv}
\end{equation}
we find
\begin{equation}
e^{i\theta}F
=
iX-Y,\nn
\end{equation}
with
\begin{equation}
X
=
-\frac{1}{2\pi r}
+
b r
+
a r\cos(2\theta),
\qquad
Y
=
a r\sin(2\theta).
\label{eq:XYdef}
\end{equation}
Thus the $b$ dependent correction is isotropic, whereas the $a$ correction is anisotropic and explicitly depends on the orientation of the relative coordinate. For equal like-signed vortices, $\Gamma_1=\Gamma_2=\Gamma>0$, the dissipative relative equation is
\begin{equation}
\frac{d\overline w}{dt}
=
2\Gamma(1+i\gamma)F(w,\overline w).\nn
\end{equation}
Using Eq.~\eqref{eq:XYdef}, the leading torus-corrected equations become
\begin{equation}
\dot r
=
\frac{\Gamma\gamma}{\pi r}
-2\Gamma a r\sin(2\theta)
-2\Gamma\gamma b r
-2\Gamma\gamma a r\cos(2\theta),
\label{eq:samesign_rdot}
\end{equation}
and
\begin{equation}
\dot\theta
=
\frac{\Gamma}{\pi r^2}
-2\Gamma b
-2\Gamma a\cos(2\theta)
+2\Gamma\gamma a\sin(2\theta).
\label{eq:samesign_thetadot}
\end{equation}
Equivalently,
\begin{equation}
\dot R
=
\frac{2\Gamma\gamma}{\pi}
-4\Gamma aR\sin(2\theta)
-4\Gamma\gamma bR
-4\Gamma\gamma aR\cos(2\theta).
\label{eq:samesign_Rdot}
\end{equation}
Hence the planar outward spiral is modified by an isotropic finite-size drift and by orientation-dependent oscillatory corrections.\\
On the other hand, for an equal opposite-sign dipole, $\Gamma_1=\Gamma$ and $\Gamma_2=-\Gamma$, the conservative relative velocity cancels, while the dissipative part gives
\begin{equation}
\frac{d\overline w}{dt}
=
-2i\gamma\Gamma F(w,\overline w).
\end{equation}
The corresponding reduced equations in polar variables are
\begin{equation}
\dot r
=
-\frac{\Gamma\gamma}{\pi r}
+
2\Gamma\gamma b r
+
2\Gamma\gamma a r\cos(2\theta),
\label{eq:dipole_rdot}
\end{equation}
and
\begin{equation}
\dot\theta
=
-2\Gamma\gamma a\sin(2\theta),
\label{eq:dipole_thetadot}
\end{equation}
or, in terms of $R=r^2$,
\begin{equation}
\dot R
=
-\frac{2\Gamma\gamma}{\pi}
+
4\Gamma\gamma bR
+
4\Gamma\gamma aR\cos(2\theta).
\label{eq:dipole_Rdot}
\end{equation}
Thus the planar dipole collapse law is recovered when $a,b\to0$, but the compact geometry changes dipole orientation.  In particular, Eq.~\eqref{eq:dipole_thetadot} shows that the anisotropic torus correction produces a slow angular drift except along the symmetry directions $\sin(2\theta)=0$.  These reduced equations provide the starting point for the explicit perturbative solutions collected below.
\subsection{Perturbative solutions for localized  dissipative binaries}
\label{perts}
We solve the local equations perturbatively by exploiting the scale separation between the singular planar interaction and the subleading finite-size corrections induced by the torus geometry. The expansion is controlled by the small-separation limit $|w|\ll 1$, in which the leading term $-i/(2\pi w)$ dominates over the regular contributions proportional to $a(\rho)w$ and $b(\rho)\overline{w}$. At zeroth order, the dynamics therefore reduces to the universal planar dissipative problem, whose closed-form solutions are summarized in Sec.~\ref{2vdiss}. The leading corrections are obtained by evaluating the $a$- and $b$-dependent terms on the planar solution $(R_0(t),\theta_0(t))$ and retaining contributions linear in $a$ and $b$. In particular, the time dependence of $\theta(t)$ entering through the anisotropic $a$-term feeds back into the radial dynamics only at $O(a^2)$, and can therefore be consistently approximated by $\theta_0(t)$ at first order. This procedure yields explicit analytic corrections to both $R(t)$ and $\theta(t)$, capturing the leading geometric effects of the torus while preserving the integrable structure of the local planar dynamics. The accuracy of this perturbative description is verified numerically in Appendix~\ref{app3}. Here we collect the explicit analytic solutions for the dissipative two-vortex problem on a doubly periodic fluid domain, including both the leading planar contributions and the first torus corrections arising from the local expansion of the interaction kernel,
\begin{equation}
F(w,\overline w)
=
-\frac{i}{2\pi w}
+i a w
+i b\overline w
+\cdots,
\end{equation}
where $a$ and $b$ are real geometry-dependent coefficients determined by the torus modular parameter. We present the solution in relative variables $w = r e^{i\theta},$ and $R = r^2$
and denote the initial conditions by
\begin{equation}
R(0)=R_i,
\qquad
\theta(0)=\theta_i.
\end{equation}
For equal like-signed vortices $\Gamma_1=\Gamma_2=\Gamma>0$, the planar dissipative solution (i.e., the zeroth-order solution in the perturbative scheme) is given by the logarithmic spiral derived in Sec.~\ref{2vdiss} (note that the variable $R = r^2$)
\begin{equation}
R_0(t)
=
R_i + \frac{2\Gamma\gamma}{\pi}t,
\end{equation}
\begin{equation}
\theta_0(t)
=
\theta_i
+
\frac{1}{2\gamma}
\log\!\left(\frac{R_0(t)}{R_i}\right).
\end{equation}
The torus-corrected radial solution to first order in $a$ and $b$ is
\begin{align}
R(t)
&=
R_0(t)
-\pi b\left[R_0^2(t)-R_i^2\right]
\nonumber\\
&\quad
-\frac{2\pi a}{\gamma}
\left[\mathcal S(R_0(t))-\mathcal S(R_i)\right]
-2\pi a
\left[\mathcal C(R_0(t))-\mathcal C(R_i)\right],
\end{align}
where
\begin{equation}
\Phi(R)
=
2\theta_i + \frac{1}{\gamma}\log\!\left(\frac{R}{R_i}\right),
\end{equation}
and
\begin{equation}
\mathcal S(R)
=
\frac{R^2}{4+\gamma^{-2}}
\left[
2\sin\Phi(R)
-
\frac{1}{\gamma}\cos\Phi(R)
\right],
\end{equation}
\begin{equation}
\mathcal C(R)
=
\frac{R^2}{4+\gamma^{-2}}
\left[
2\cos\Phi(R)
+
\frac{1}{\gamma}\sin\Phi(R)
\right].
\end{equation}
The corresponding angular solution is obtained by linearization,
\begin{equation}
\theta(t)
=
\theta_0(t)
+
\int_0^t
\left[
-\frac{\Gamma}{\pi R_0^2(s)}\bigl(R(s)-R_0(s)\bigr)
-2\Gamma b
-2\Gamma a\cos(2\theta_0(s))
+2\Gamma\gamma a\sin(2\theta_0(s))
\right]ds.
\end{equation}
The angular correction for the same-sign binary can be evaluated in closed form. Writing the full angular solution as
\begin{equation}
\theta(t)
=
\theta_0(t)
+
\theta_{\mathrm{corr}}(t),
\end{equation}
the correction admits the explicit closed-form expression 
\begin{align}
\theta_{\mathrm{corr}}(t)
&=
\frac{1}{(1+4\gamma^2)\,R_0(t)}
\Bigg\{
-2\Gamma\gamma b\,
\bigl(R_0(t)-R_i\bigr)
\nonumber\\
&\quad
+
a\,\mathcal{F}_c(t)\cos(2\theta_i)
+
a\,\mathcal{F}_s(t)\sin(2\theta_i)
\Bigg\}.
\label{eq:thetaCorr_closed}
\end{align}
The functions $\mathcal{F}_c(t)$ and $\mathcal{F}_s(t)$ collect the oscillatory contributions,
\begin{align}
\mathcal{F}_c(t)
&=
\left(\pi R_i+2\Gamma\gamma t\right)^2
\cos\!\bigl(\Phi(t)\bigr)
+
\pi R_i\left(3\gamma \pi R_i + \Gamma t\right)
\nonumber\\
&\quad
-
\left(\pi R_i+2\Gamma\gamma t\right)^2
\sin\!\bigl(\Phi(t)\bigr),
\\
\mathcal{F}_s(t)
&=
\left(\pi R_i+2\Gamma\gamma t\right)^2
\sin\!\bigl(\Phi(t)\bigr)
+
\pi R_i\left(\gamma \pi R_i + \Gamma t\right)
\nonumber\\
&\quad
+
\left(\pi R_i+2\Gamma\gamma t\right)^2
\cos\!\bigl(\Phi(t)\bigr).
\end{align}
Thus, the full solution may be expressed compactly as
\begin{equation}
\theta(t)
=
\theta_i
+
\frac{1}{2\gamma}
\log\!\left(\frac{R_0(t)}{R_i}\right)
+
\theta_{\mathrm{corr}}(t),
\end{equation}
with $\theta_{\mathrm{corr}}(t)$ given by Eq.~\eqref{eq:thetaCorr_closed}.\\
For an equal opposite-sign dipole, $\Gamma_1=\Gamma$ and $\Gamma_2=-\Gamma$, the planar solution (ie. the zeroth order solution) describes linear collapse at fixed orientation, as found in Sec.~\ref{2vdiss} (note again that the variable $R = r^2$)
\begin{equation}
R_0(t)
=
R_i - \frac{2\Gamma\gamma}{\pi}t,
\qquad
\theta_0(t)=\theta_i.
\end{equation}
The torus-corrected angular dynamics is autonomous and admits the exact solution
\begin{equation}
\tan\theta(t)
=
\tan\theta_i\,e^{-4\Gamma\gamma a t},
\end{equation}
which yields
\begin{equation}
\theta(t)
=
\tan^{-1}\!\left[
\tan\theta_i\,e^{-4\Gamma\gamma a t}
\right],
\end{equation}
For the radial solution, it is helpful to introduce the notation
\begin{equation}
q(t)=\tan\theta(t),
\qquad
q_i=\tan\theta_i.
\end{equation}
The radial equation becomes linear and can be written in closed form as
\begin{equation}
R(t)
=
e^{M(t)}
\left[
R_i
-
\frac{2\Gamma\gamma}{\pi}
\int_0^t e^{-M(s)}\,ds
\right],
\end{equation}
with exponent
\begin{equation}
M(t)
=
4\Gamma\gamma(a+b)t
+
\log\!\left(\frac{1+q^2(t)}{1+q_i^2}\right).
\end{equation}
For generic $a\neq b$ and $q_i\neq 0$, the integral can also be written in closed form in terms of Gauss hypergeometric functions (as obtained using \textit{Mathematica}),
\begin{align}
R(t)
&=
\frac{
e^{4\Gamma\gamma(a+b)t}
\left(1+q_i^2 e^{-8a\Gamma\gamma t}\right)
}{
1+q_i^2
}
\Bigg[
R_i
+
\frac{1+q_i^2}{2\pi (a-b)q_i^2}
\Bigg\{
{}_2F_1\!\left(
1,\frac{a-b}{2a};
\frac{3}{2}-\frac{b}{2a};
-\frac{1}{q_i^2}
\right)
\nonumber\\
&\hspace{4.5cm}
-
e^{-4\Gamma\gamma(a-b)t}
{}_2F_1\!\left(
1,\frac{a-b}{2a};
\frac{3}{2}-\frac{b}{2a};
-\frac{e^{8a\Gamma\gamma t}}{q_i^2}
\right)
\Bigg\}
\Bigg].
\end{align}
where $q_i=\tan\theta_i,
q(t)=q_i e^{-4\Gamma\gamma a t}$.\\
To leading roder, the solution can be written in a simpler form
\begin{equation}
R(t)
=
R_i
-\frac{2\Gamma\gamma}{\pi}t
+
4\Gamma\gamma
\left[
b+a\cos(2\theta_i)
\right]
\left[
R_i t
-
\frac{\Gamma\gamma}{\pi}t^2
\right],
\end{equation}
and
\begin{equation}
\theta(t)
=
\theta_i
-
2\Gamma\gamma a\,t\sin(2\theta_i)
+
O(a^2).
\end{equation}
Importantly, for the equal dipole, the planar fixed-orientation is lost and
 the compact geometry induces a slow orientation drift. Thus, even at small separation, the compact flat domain ie. the flat torus leaves a measurable imprint on dissipative binary motion through both radial and angular corrections. In both cases, the planar dynamics is recovered in the limit $a,b\to0$. The torus corrections introduce geometry-dependent secular and oscillatory contributions, which are essential for quantitative agreement with the full numerical dynamics, as demonstrated in the accompanying numerical Appendix \ref{app3}.
\section{Conclusion}
\label{cncl}
We have developed a reduced theory of dissipative point-vortex dynamics on a doubly periodic fluid domain (flat torus), using the exact Hamiltonian formulation in terms of the Schottky-Klein prime function as the conservative problem. The conservative dynamics is governed by a torus interaction kernel \(F\) satisfying the antisymmetry relation \(F(1/\zeta)=-F(\zeta)\). This identity implies exact solvability of the binary problem in terms of the relative coordinate and gives a starting point for comparing same-sign pairs, unequal binaries, and vortex dipoles. In particular, an equal opposite-sign pair has vanishing total circulation and therefore translates rigidly with fixed separation in the nondissipative problem, while equal same-sign vortices execute relative orbital motion determined by the real and imaginary parts of the torus kernel.

We then introduced a minimal dissipative correction by adding to each Hamiltonian vortex velocity a component rotated by \(90^\circ\), as motivated by mutual-friction phenomenology in finite-temperature superfluids and dissipative vortex models of Bose-Einstein condensates. This converts the conservative dynamics into a mixed symplectic-gradient flow. For quantized vortices with circulation sign identified with the vortex orientation, the Hamiltonian obeys
\[
\frac{dH}{dt}
=
-4\gamma
\sum_j
\left|
\frac{\partial H}{\partial w_j}
\right|^2
\le 0,
\]
so the dissipative dynamics is energy decreasing.  For the vortex binary, the dissipative equations remain analytically tractable in the local regime. The singular planar part of the torus kernel gives simple closed-form laws. Equal same-sign vortices form an outward spiral,
\[
r^2(t)=r_0^2+\frac{2\Gamma\gamma}{\pi}t,
\]
whereas an equal opposite-sign dipole collapses at fixed orientation,
\[
r^2(t)=r_0^2-\frac{2\Gamma\gamma}{\pi}t .
\]
For an unequal opposite-sign pair, dissipation produces simultaneous contraction and rotation. Writing
\[
\Sigma=\Gamma_a+\Gamma_b,
\qquad
\Delta=\Gamma_a-\Gamma_b,
\]
we obtained
\[
r^2(t)=r_0^2-\frac{\gamma\Sigma}{\pi}t,
\qquad
\omega(t)=\frac{\Delta}{2\pi r^2(t)}.
\]
Thus the instantaneous angular frequency diverges as
\[
\omega(t)=\frac{\omega_0}{1-t/t_*},
\qquad
t_*=\frac{\pi r_0^2}{\gamma\Sigma},
\]
and satisfies the nonlinear chirp law
\[
\dot\omega
=
\frac{2\gamma\Sigma}{\Delta}\omega^2.
\]
This provides a simple hydrodynamic example of finite-time frequency growth generated by dissipative collapse. Finally, we derived the leading corrections due to the compact torus geometry. Expanding the interaction kernel as
\[
F(w,\overline w)
=
-\frac{i}{2\pi w}
+i a(\rho) w
+i b(\rho)\overline w+\cdots ,
\]
we found that \(b(\rho)\) produces isotropic secular corrections, while \(a(\rho)\) produces anisotropic angular modulation. For same-sign vortices, the outward spiral persists but acquires geometry-dependent secular and oscillatory corrections. For the equal dipole, the planar fixed-orientation is lost: the angular equation becomes
\[
\dot\theta
=
-2\Gamma\gamma a(\rho)\sin(2\theta),
\]
so the compact geometry induces a slow orientational drift, except along symmetry directions. \textit{Even at small separations, the compact fluid domain geometry leaves a measurable imprint on dissipative binary dynamics through both radial and angular corrections}. This constitutes a central result of the present study.\\
These results show that dissipation and compact geometry combine in a nontrivial way. The conservative torus problem supplies an exactly integrable binary reference system, while the dissipative extension produces monotonic energy decay, outward spiraling of same-sign pairs, collapse of dipoles, and chirping unequal binaries. The local expansion further reveals how the global periodic geometry enters through a small number of coefficients that modify the planar laws. This framework should be useful for interpreting vortex motion in finite and periodic superfluid systems, and it provides a basis for extending the analysis to many-vortex relaxation, metastable vortex clusters, and geometry-dependent dissipative transport on compact domains.

\section{Acknowledgments}
It is a pleasure to thank Takashi Sakajo, Suryateja Gavva, Naomi Oppenheimer, Haim Diamant. R.S is supported by DST INSPIRE Faculty fellowship, India (Grant No.IFA19-PH231). R.S. acknowledges support from NFSG and OPERA Research Grant from Birla Institute of Technology and Science, Pilani (Hyderabad Campus).
\section{Data Availability}
The data supporting the findings of this study are available within the article. Numerical implementations are available from the corresponding author upon reasonable request.

\appendix

\section{ Notes on the Schottky-Klein formulation }
\label{app1}
Vortex interactions on doubly periodic domains can be expressed in terms of the Schottky-Klein prime function, a special function defined on multiply connected circular regions~\cite{Crowdy2005}. For the annular domain
\[
D_\zeta = \{\zeta \in \mathbb{C} \;|\; \rho < |\zeta| \le 1\}, \qquad 0<\rho<1,
\]
the prime function admits the infinite-product representation
\begin{equation}
P(\zeta,\sqrt{\rho})
=
(1-\zeta)
\prod_{k=1}^{\infty}
(1-\rho^k\zeta)(1-\rho^k/\zeta).
\label{eq:Pfunction_new}
\end{equation}
This function has a simple zero at $\zeta=1$ and encodes the periodic image structure of the domain. The quantity entering the vortex dynamics is the logarithmic derivative of $P$, defined as
\begin{equation}
K(\zeta,\sqrt{\rho})
=
\frac{\zeta\,\partial_\zeta P(\zeta,\sqrt{\rho})}{P(\zeta,\sqrt{\rho})},
\label{eq:Kdef_new}
\end{equation}
where $\partial_\zeta$ denotes differentiation with respect to $\zeta$. Expanding Eq.~\eqref{eq:Pfunction_new} yields the convergent series
\begin{align}
K(\zeta,\sqrt{\rho})
&=
\frac{\zeta}{\zeta-1}
+
\sum_{k=1}^{\infty}
\left[
\frac{-\rho^k\zeta}{1-\rho^k\zeta}
+
\frac{\rho^k/\zeta}{1-\rho^k/\zeta}
\right].
\label{eq:Kseries_new}
\end{align}
In particular, near the singular point $\zeta=1$,
\begin{equation}
K(\zeta,\sqrt{\rho})
=
\frac{1}{\zeta-1}+O(1),
\qquad \zeta\to 1,\nn
\end{equation}
so the leading behaviour coincides with the planar interaction, with periodic corrections entering at subleading order. A closed representation of $K$ in terms of $q$-special functions was obtained in~\cite{sam3}:
\begin{equation}
K(\zeta,\sqrt{\rho})
=
\frac{1}{1-\zeta}
+
\frac{1}{\log\rho}
\left[
\psi_\rho\!\left(\frac{\log(1/\zeta)}{\log\rho}\right)
-
\psi_\rho\!\left(\frac{\log\zeta}{\log\rho}\right)
\right],
\label{eq:K_qdigamma_new}
\end{equation}
where $\psi_\rho$ is the $q$-digamma function with base $q=\rho$ determined by  the geometry of the torus. It is defined as the logarithmic derivative of the $q$-gamma function $\Gamma_\rho(z)$,
\begin{equation}
\psi_\rho(z)=\frac{d}{dz}\log \Gamma_\rho(z),
\qquad
\Gamma_\rho(z)
=
(1-\rho)^{1-z}
\prod_{n=0}^{\infty}
\frac{1-\rho^{n+1}}{1-\rho^{n+z}},
\label{eq:qgamma_new}
\end{equation}
which converges for $0<\rho<1$ and complex $z$.
The structure of $\Gamma_\rho$ reveals the periodic nature of the geometry. Its poles arise from zeros of the denominator factors $(1-\rho^{n+z})$, which occur when
\begin{equation}
\rho^{\,n+z}=1
\quad\Rightarrow\quad
(n+z)\ln\rho=2\pi i k,
\qquad k\in\mathbb{Z}.\nn
\end{equation}
Solving for $z$ gives
\begin{equation}
z_{n,k}
=
-\,n+\frac{2\pi i k}{\ln\rho},
\qquad n,k\in\mathbb{Z}.\nn
\end{equation}
Thus the poles form a doubly periodic lattice in the complex $z$-plane, with unit spacing along the real axis and vertical spacing $2\pi/|\ln\rho|$, reflecting the underlying periodic image structure of the flat torus).

\section{Notes on the local expansion of the kernel $K(z)$}
\label{app2}
We obtain the small-$z$ expansion of the interaction kernel
\begin{equation}
K(z)
=
\frac{1}{1-e^z}
+
\frac{1}{\log\rho}
\left[
\psi_{\rho}\!\left(-\frac{z}{\log\rho}\right)
-
\psi_{\rho}\!\left(\frac{z}{\log\rho}\right)
\right],
\qquad 0<\rho<1,\nn
\end{equation}
where $\psi_\rho$ is the $q$-digamma function and $\log\rho<0$. For notational convenience, we define
\begin{equation}
L=\log\rho, \qquad x=\frac{z}{L},\nn
\end{equation}
so that
\begin{equation}
K(z)
=
\frac{1}{1-e^z}
+
\frac{1}{L}\bigl[\psi_\rho(-x)-\psi_\rho(x)\bigr].\nn
\end{equation}
We first isolate the elementary contribution. Expanding about $z=0$ gives
\begin{equation}
\frac{1}{1-e^z}
=
-\frac{1}{z}
+\frac{1}{2}
-\frac{z}{12}
+\frac{z^3}{720}
+O(z^5).\nn
\end{equation}
To treat the $q$-digamma term, we make use of the shift relation (valid for $0<\rho<1$)
\begin{equation}
\psi_\rho(u+1)-\psi_\rho(u)
=
-\frac{(\log\rho)\,\rho^u}{1-\rho^u},\nn
\end{equation}
which may be rearranged as
\begin{equation}
\psi_\rho(u)
=
\psi_\rho(1+u)
+
(\log\rho)\frac{\rho^u}{1-\rho^u}.\nn
\end{equation}
Applying this identity to $u=\pm x$ yields
\begin{equation}
\psi_\rho(x)
=
\psi_\rho(1+x)
+
L\frac{e^{Lx}}{1-e^{Lx}},
\qquad
\psi_\rho(-x)
=
\psi_\rho(1-x)
+
L\frac{e^{-Lx}}{1-e^{-Lx}}.\nn
\end{equation}
Taking the difference separates the analytic and singular parts:
\begin{equation}
\psi_\rho(-x)-\psi_\rho(x)
=
\bigl[\psi_\rho(1-x)-\psi_\rho(1+x)\bigr]
+
L\left[
\frac{e^{-Lx}}{1-e^{-Lx}}
-
\frac{e^{Lx}}{1-e^{Lx}}
\right].\nn
\end{equation}
We now expand each contribution for small $x$. The regular part is obtained by Taylor expanding about $x=0$:
\begin{equation}
\psi_\rho(1\pm x)
=
\psi_\rho(1)
\pm \psi_\rho^{(1)}(1)x
+O(x^2),\nn
\end{equation}
which gives
\begin{equation}
\psi_\rho(1-x)-\psi_\rho(1+x)
=
-2\psi_\rho^{(1)}(1)x
+O(x^3).\nn
\end{equation}
The rational terms are expanded using
\begin{equation}
\frac{e^u}{1-e^u}
=
-\frac{1}{u}
-\frac{1}{2}
-\frac{u}{12}
+O(u^3),
\qquad
\frac{e^{-u}}{1-e^{-u}}
=
\frac{1}{u}
-\frac{1}{2}
+\frac{u}{12}
+O(u^3).\nn
\end{equation}
Substituting $u=Lx$ gives
\begin{equation}
L\frac{e^{Lx}}{1-e^{Lx}}
=
-\frac{1}{x}
-\frac{L}{2}
-\frac{L^2}{12}x
+O(x^3),\nn
\end{equation}
\begin{equation}
L\frac{e^{-Lx}}{1-e^{-Lx}}
=
\frac{1}{x}
-\frac{L}{2}
+\frac{L^2}{12}x
+O(x^3). \nn
\end{equation}
Their difference therefore reads
\begin{equation}
L\left[
\frac{e^{-Lx}}{1-e^{-Lx}}
-
\frac{e^{Lx}}{1-e^{Lx}}
\right]
=
\frac{2}{x}
+
\frac{L^2}{6}x
+O(x^3). \nn
\end{equation}
Combining the analytic and singular pieces, we obtain
\begin{equation}
\psi_\rho(-x)-\psi_\rho(x)
=
\frac{2}{x}
+
\left(
\frac{L^2}{6}
-
2\psi_\rho^{(1)}(1)
\right)x
+O(x^3).\nn
\end{equation}
Dividing by $L$ and restoring $x=z/L$ gives
\begin{equation}
\frac{1}{L}\bigl[\psi_\rho(-x)-\psi_\rho(x)\bigr]
=
\frac{2}{z}
+
\left(
\frac{1}{6}
-
\frac{2\psi_\rho^{(1)}(1)}{L^2}
\right)z
+O(z^3).\nn
\end{equation}
Finally, adding the elementary contribution yields
\begin{align}
K(z)
&=
\left(
-\frac{1}{z}
+\frac{1}{2}
-\frac{z}{12}
\right)
+
\left(
\frac{2}{z}
+
\left(
\frac{1}{6}
-
\frac{2\psi_\rho^{(1)}(1)}{L^2}
\right)z
\right)
+O(z^3)\nn
\\
&=
\frac{1}{z}
+
\frac{1}{2}
+
\left(
\frac{1}{12}
-
\frac{2\psi_\rho^{(1)}(1)}{\log^2\rho}
\right)z
+O(z^3). \nn
\end{align}
in agreement with the expression used in the main text.

\section{Numerical verification of dissipative binary dynamics on the torus}
\label{app3}
In this appendix we present a detailed numerical verification of the analytic results derived in the main text for dissipative vortex binaries on a doubly periodic domain. We consider both equal opposite-sign vortices (dipole) and equal same-sign vortices, and compare the full numerical solution of the equations of motion with the leading planar theory and the torus-corrected analytic solutions obtained from the local expansion of the interaction kernel.

All simulations are performed on a flat torus characterized by the modular parameter
\begin{equation}
\rho = e^{-\pi},
\end{equation}
with domain extents $x\in[0,2\pi]$ and $y\in[0,-\log\rho]$. The dissipative parameter is fixed to
\begin{equation}
\gamma = 0.02,
\end{equation}
and the interaction kernel is evaluated using the Schottky-Klein prime function and its associated $q$-representation. The torus-correction coefficients entering the local expansion,
\begin{equation}
F(w,\overline w) = -\frac{i}{2\pi w} + i a w + i b \overline w + \cdots,
\end{equation}
are computed numerically for this geometry, yielding
\begin{equation}
a \approx -0.0277169,
\qquad
b \approx 0.0253303.
\end{equation}
The initial conditions are chosen as two vortices placed close to one another,
\begin{equation}
(x_1,y_1) = (\pi,\pi/2),
\qquad
(x_2,y_2) = (\pi+0.1,\pi/2+0.1),
\end{equation}
corresponding to an initial squared separation
\begin{equation}
R_i \approx 0.02,
\end{equation}
and an initial relative angle
\begin{equation}
\theta_i \approx -2.35619.
\end{equation}
The numerical trajectories are obtained by solving the full vortex equations of motion and extracting the relative variables
\begin{equation}
R(t) = |\mathbf r_1 - \mathbf r_2|^2,
\qquad
\theta(t) = \arg(\mathbf r_1 - \mathbf r_2).
\end{equation}
We first consider the equal opposite-sign dipole case with circulations $\Gamma_1=1$ and $\Gamma_2=-1$. In the planar limit, the dynamics predicts a purely radial collapse at fixed orientation,
\begin{equation}
R(t) = R_i - \frac{2\Gamma\gamma}{\pi}t,
\qquad
\theta(t)=\theta_i,
\end{equation}
with collapse time $t_c = \pi R_i/(2\Gamma\gamma)$. The torus-corrected solution modifies both radial and angular dynamics through the geometry-dependent coefficients $a$ and $b$, leading to an exponential relaxation of the orientation and a corrected radial evolution obtained from the integrating factor solution derived in the main text.

\begin{figure}[t]
\centering
\includegraphics[width=\linewidth]{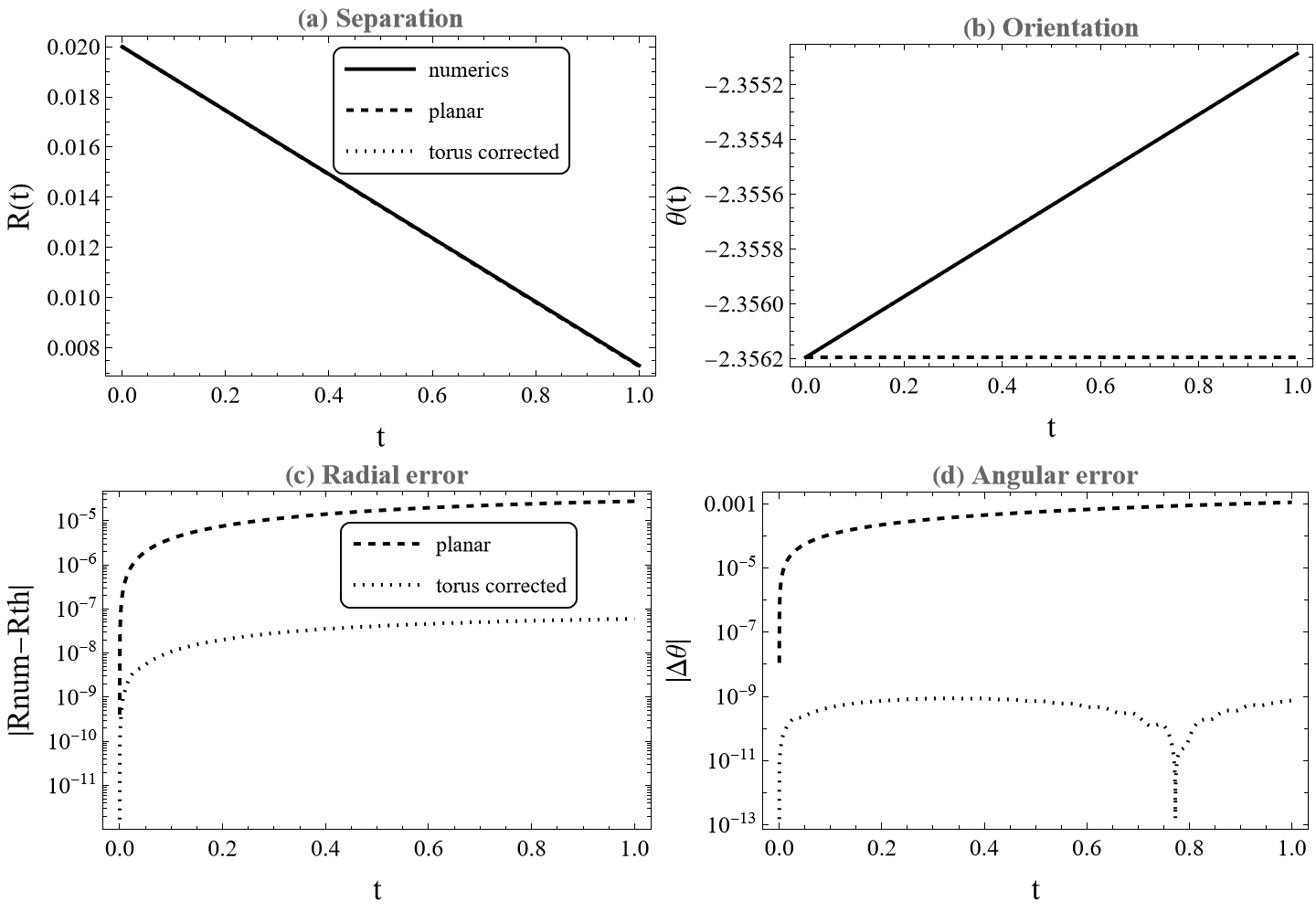}
\caption{
Numerical verification of the dissipative dipole dynamics ($\Gamma_1=1$, $\Gamma_2=-1$) on a flat torus.
(a) Separation $R(t)$ showing agreement between numerics (solid), planar theory (dashed), and torus-corrected solution (dotted).
(b) Orientation $\theta(t)$, where the planar theory predicts no evolution, while the torus correction captures the observed drift.
(c,d) Radial and angular errors. The torus-corrected solution reduces the error by several orders of magnitude compared to the planar approximation. Sharp minima correspond to zero-crossings of the error and do not indicate improved accuracy.
}
\label{fig:dipole_verification}
\end{figure}

Figure~\ref{fig:dipole_verification} shows the comparison between numerics, planar theory, and torus-corrected theory. The separation $R(t)$ is accurately captured by both theories at leading order, but the torus correction significantly improves the agreement, especially in the angular dynamics where the planar theory predicts no evolution. The error plots demonstrate that the torus-corrected solution reduces the discrepancy by several orders of magnitude in both $R(t)$ and $\theta(t)$ over the full time interval prior to collapse.

We next consider the equal same-sign case with circulations $\Gamma_1=\Gamma_2=1$. In the planar limit, the pair executes an outward logarithmic spiral,
\begin{equation}
R(t) = R_i + \frac{2\Gamma\gamma}{\pi}t,
\qquad
\theta(t) = \theta_i + \frac{1}{2\gamma}\log\left(\frac{R(t)}{R_i}\right),
\end{equation}
while the torus corrections introduce both secular and oscillatory modifications to the radial growth as well as corrections to the angular evolution through the geometry-dependent terms proportional to $a$ and $b$.

\begin{figure}[t]
\centering
\includegraphics[width=.8\linewidth]{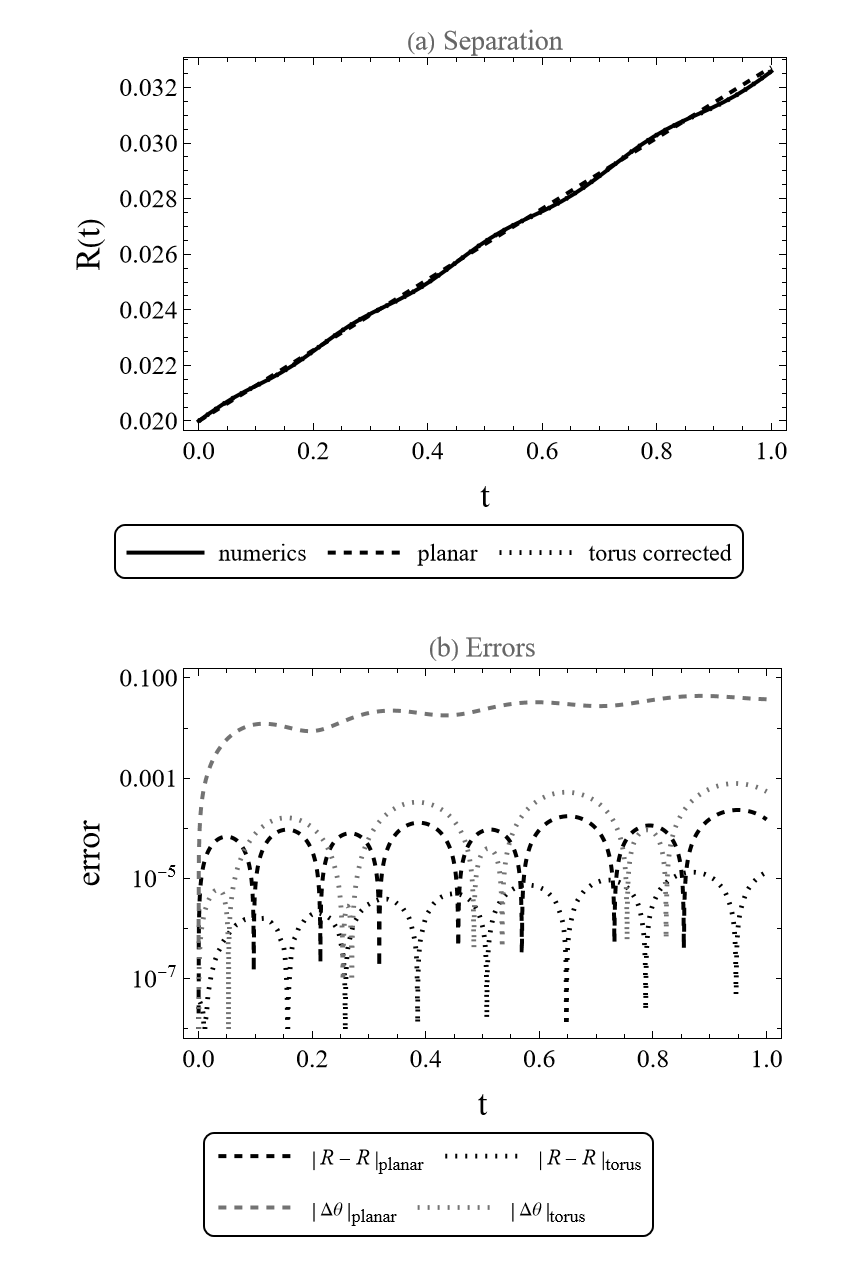}
\caption{Numerical verification of the dissipative same-sign vortex pair ($\Gamma_1=\Gamma_2=1$). (a) Separation $R(t)$ showing outward growth and improved agreement of the torus-corrected solution with numerics. (b) Combined radial and angular error comparison. The torus-corrected theory significantly reduces both errors relative to the planar logarithmic-spiral solution. The oscillatory structure reflects the phase-dependent corrections along the trajectory, while the overall reduction confirms the importance of toroidal geometry.}
\label{fig:samesign_verification}
\end{figure}

The numerical comparison is shown in Fig.~\ref{fig:samesign_verification}. The torus-corrected solution closely tracks the numerical trajectory, while the planar approximation exhibits systematic deviations that grow in time. The error analysis confirms that the inclusion of torus corrections reduces both radial and angular errors by one to two orders of magnitude. The oscillatory structure observed in the error curves arises from the phase-dependent nature of the correction terms and the logarithmic winding of the trajectory; sharp dips correspond to isolated zero-crossings of the error. Overall, these numerical investigations provide strong quantitative validation of the analytic torus-corrected solutions derived in the main text. They demonstrate that while the leading singular term captures the dominant dynamics, the geometry-dependent corrections are essential for quantitative agreement and become increasingly important over longer times.

\end{document}